\documentclass[lettersize,journal]{IEEEtran}
\usepackage{amsmath,amsfonts}
\usepackage{optidef}
\usepackage{algorithmic}
\usepackage[ruled,vlined]{algorithm2e}
\include{pythonlisting}
\usepackage{array}
\usepackage[caption=false,font=normalsize,labelfont=sf,textfont=sf]{subfig}
\usepackage{textcomp}
\usepackage{stfloats}
\usepackage{color}
\usepackage{url}
\usepackage{verbatim}
\usepackage{graphicx}
\graphicspath{ {./images/} }
\usepackage{lipsum}

\usepackage{cite}
\renewcommand{\baselinestretch}{1.08}

\begin{document}

\title{A Genetic Algorithm-Based Approach to Power Allocation in Rate-Splitting Multiple Access Systems}

\author{Temitope O. Fajemilehin, Kobi Cohen \emph{(Senior Member, IEEE)}
\thanks{Temitope O. Fajemilehin and Kobi Cohen are with the School of Electrical and Computer Engineering, Ben-Gurion University of the Negev, Beer Sheva 8410501 Israel. Email: temitope@post.bgu.ac.il, yakovsec@bgu.ac.il}
\thanks{This work was supported by the Israel Science Foundation under Grant 2640/20.}
\thanks{This work has been submitted to the IEEE for possible publication. Copyright may be transferred without notice, after which this version may no longer be accessible.}
}

\maketitle

\begin{abstract}
We consider the problem of power allocation in Rate-Splitting Multiple Access (RSMA) systems, where messages are split into common and private messages. The common and private streams are jointly transmitted to allow efficient use of the bandwidth, and decoded by Successive Interference Cancellation (SIC) at the receiver. However, the power allocation between streams significantly affects the overall performance. In this letter, we address this problem. We develop a novel algorithm, dubbed Power Allocation in RSMA systems using Genetic Algorithm (PARGA), to allocate the power between streams in RSMA systems in order to maximize the user sum-rate. Simulation results demonstrate the high efficiency of PARGA compared to existing methods.  
\end{abstract}


\section{Introduction}

The continuously growing need for wireless communication services has driven the telecommunications research to devise cutting-edge algorithms and solutions aimed at enhancing performance. As the world becomes increasingly connected, users expect high-quality wireless experiences for a multitude of applications, from ultra-fast internet browsing to real-time video streaming and IoT connectivity. Meeting these escalating demands presents a significant challenge, one that hinges on both bandwidth efficiency and power allocation. In this context, the efficient utilization of limited spectral resources becomes paramount, as does the efficient allocation of transmit power to maximize system performance. This paper focuses on multiple access schemes, aimed at enhancing both bandwidth efficiency and power utilization in wireless networks.

Multiple access schemes are used to allow multiple users to share the radio spectrum efficiently. A variety of multiple access methods have been developed in the past and more recently to support the increased demand of services in wireless communication networks. Orthogonal multiple access (OMA) techniques have played a great role in traditional communication networks. In OMA techniques, users transmit data using orthogonal channels across frequency and time (e.g., OFDMA), motivating extensive research on resource allocation over orthogonal channels \cite{zhang2012resource, cohen2013game, di2016joint,  cohen2016distributed, naparstek2018deep, dehghani2018joint,   gafni2020learning, jain2020adaptive,  gafni2022distributed}. However, OMA methods suffer from a limited connectivity under diverse quality of services (QoS) demands of users, fairness guarantees, and scarce spectral resources (see \cite{alghasmari2020power, sery2020analog, sery2021over, paul2021accelerated, gez2023subgradient} and references therein). A more recent multiple access scheme is the Space Division Multiple Access (SDMA) method. SDMA separates users spatially and superposes them using the same frequency and time \cite{sharma2016multiple}. It uses linear precoding to distinguish users and treat any residual interference as noise. Another multiple access scheme is the Non-Orthogonal Multiple Access (NOMA), that superposes users in the same frequency and time while separating them based on the power or code domain \cite{islam2016power, wei2016survey, akbar2021noma, budhiraja2021systematic}. In NOMA, users are superposed using linearly precoded superposition coding and decoded by successive interference cancellation (SIC), allowing for full interference decoding, unlike SDMA.

RSMA is a new multiple access technique that improves the spectrum and energy efficiency by generalizing SDMA and NOMA \cite{mao2018rate, li2020resource, dizdar2020rate, yang2021optimization, hieu2021optimal, mao2022rate, huang2022deep, chopra2023review, tran2023review}. RSMA splits each user’s message into a common part and a private part. This is done with the aid of linearly precoded rate-splitting (RS). The common message stream and the linearly precoded private message streams are jointly transmitted. At the receiver, the common message stream is decoded by each user and then the common message stream is removed using SIC before the private stream is decoded. 

Even though RSMA has demonstrated improved performance compared to SDMA and NOMA, the allocation of the total transmitted power between the common and private streams affects the optimal performance of the system. This problem is non-convex, and various methods have been proposed recently to tackle it. Reinforcement learning has been suggested in \cite{hieu2021optimal, huang2022deep}. However, it requires a lot of data for training, and the performance highly depends on the neural network architecture. In \cite{li2020resource}, the authors converted the problem into a difference of convex program, and approximated it by its first-order Taylor expansion. Then, they converted the problem into a user-subcarrier matching problem and used the Hungarian algorithm to solve it. Finally, they calculated the power allocation among the subcarriers, and updated the power vector for each user. This requires high computational complexity (cubic) due to the Hungarian method, and offers only approximate solution by first-order Taylor expansion. In \cite{yang2021optimization}, the authors suggested an iterative algorithm to achieve a sub-optimal solution.  

In this letter, we develop a novel genetic algorithm (GA)-based approach to solve the power allocation problem. GA has been used to solve the power allocation problem in downlink NOMA \cite{alghasmari2020power}. Even though the power demands in NOMA are different from RSMA, the solution produced by the GA algorithm was comparable to the full search power that gives an optimal performance, which motivates the development in this research. Specifically, we develop a novel algorithm, dubbed Power Allocation in RSMA systems using Genetic Algorithm (PARGA). PARGA formulates a GA-based optimization used to allocate the power between streams in RSMA systems in order to maximize the user sum-rate. It leverages the properties of GA, enabling it to converge to an efficient power allocation solution while efficiently handling a large search space with low complexity. Extensive simulation results that we performed demonstrate strong performance of PARGA in practical RSMA systems. 

\section{System Model and Problem Statement}

We consider a downlink communication system, comprising of a base station (BS) with \(N_t\) transmit antennas and \(K\) users, each equipped with a single antenna. The bandwidth is divided into $G$ orthogonal subcarriers. Unlike OFDMA, in RSMA each subcarrier can be shared by multiple users simultaneously. The channel state $N_t \times 1$ vector of user $k$ on channel $g$ is denoted by $\boldsymbol{h}_{k,g}$. The channel state information (CSI) is available at the BS via pilot signals, as commonly assumed in the literature \cite{mao2018rate, li2020resource, dizdar2020rate, mao2022rate, sery2020analog, sery2021over}. 

\begin{figure*}
\label{fig_1}
\includegraphics[width=\textwidth]{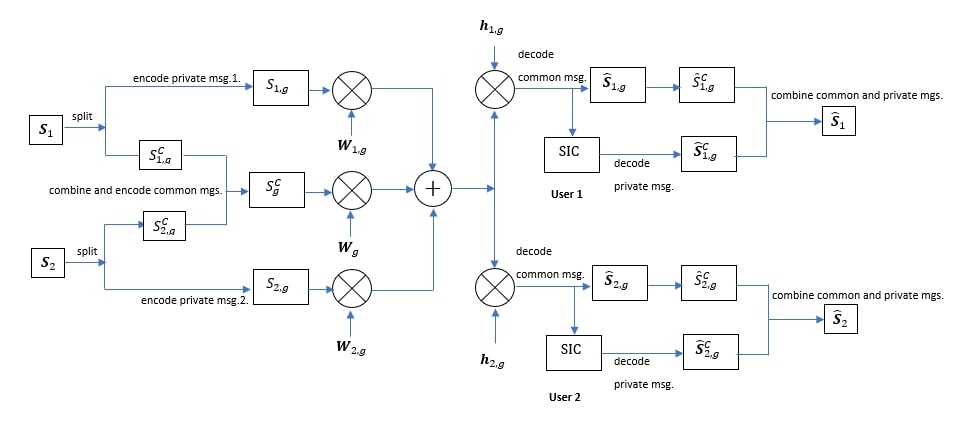}
\centering
\captionsetup{justification=centering}
\caption{An illustration of two-user RSMA transmission model on channel $g$.}
\end{figure*}

We next describe the RS transmission scheme as introduced in \cite{mao2018rate}. An illustration of a two-user RSMA transmission scheme is presented in Fig. 1. Let $\mathcal{K}_g$ be the set of users that transmit on channel $g$, with cardinality $|\mathcal{K}_g|=K_g$. Consider user $k$ that transmits its message $s_{k}$ on channel $g$. The message comprises of a common part and a private part on channel $g$, denoted by $s_{k, g}^C$ and $s_{k, g}$, respectively. The super common message $s_{g}^C$ consists of a common message split from $K_g$ pieces of common messages of the $K_g$ users on channel $g$. The messages are encoded through power domain multiplexing. The transmitted signal is given by: 
\begin{equation}
\label{eqn2}
\boldsymbol{x}_g=P_{g}^C s_{g}^C\boldsymbol{w}_{g}^C +\sum_{k\in \mathcal{K}_g} P_{k,g} s_{k,g}\boldsymbol{w}_{k,g},
\end{equation}
where $P_{g}^C, P_{k,g}$ are the power coefficients of the common message and private message of user $k$, respectively, \(\boldsymbol{w}_{g}^C\) is a random precoder for the common signal, and $\boldsymbol{w}_{k,g}$ is a ZF precoder of the private signal of user $k$,  $\boldsymbol{w}_{k,g} \subseteq \mbox{span}\Big(\{\hat{\boldsymbol{h}}^\perp_{k\prime, g}\}_{k\prime\in\mathcal{K}_g\backslash{\{k\}}}\Big)
$, for $k\in \mathcal{K}_g$. Let $P_g=P_{g}^C+\sum_{k\in\mathcal{K}_g}P_{k,g}$ denote the total allocated power at channel $g$, and let $P_T$ be the total transmitting power of the BS. Then, $\sum_{g=1}^{G}P_g\leq P_T$.

The received signal of user $k\in\mathcal{K}_g$ on channel $g$ is given by: 
\begin{equation}
\label{eqn1}
y_{k,g}=\boldsymbol{h}_{k,g}^H \boldsymbol{x}_g + n_{k,g}, 
\end{equation}
where \(n_{k,g} \sim CN(0,N_0)\) denotes the additive white Gaussian noise (AWGN) with zero mean and variance $N_0$.

Each user first decodes the common message, where the rest of the message is treated as an interference. Hence, the SINR of the common message of user $k$ on channel $g$ is given by:
\begin{equation}
\label{eqn6}
\displaystyle
\mbox{SINR}^C_{k,g} =\frac{P_{g}^C|\boldsymbol{h}_{k,g}^H \boldsymbol{w}_{g}^C|^2}{\sum_{k\in K_g} P_{k,g}|\boldsymbol{h}_{k,g}^H \boldsymbol{w}_{k,g}|^2 + N_0}
\end{equation}

The private message is decoded by successive interference cancellation (SIC). Specifically, the common message is first subtracted from $y_{k,g}$ and then the private message is decoded. Hence, the SINR of the private message of user $k$ on channel $g$ is given by:
\begin{equation}
\label{eqn7}
\mbox{SINR}_{k,g} =\frac{P_{k,g}|\boldsymbol{h}_{k,g}^H \boldsymbol{w}_{k,g}|^2}{\sum_{m\in K_g,m\neq k } P_{m,g}|\boldsymbol{h}_{k,g}^H \boldsymbol{w}_{m,g}|^2 + N_0}.
\end{equation}

To ensure that the common message is successfully decoded by all users, the rate of the common message is set to be by the Shannon rate to:
\begin{equation}
\label{eqn8}
R_{g}^C = \log_2\Big(1 + \min \{\mbox{SINR}^C_{k,g}\}\Big), k\in \mathcal{K}_g.
\end{equation}
The rate of the private message is given by:
\begin{equation}
\label{eqn9}
R_{k,g} = \log_2\Big(1 + \mbox{SINR}_{k,g}\Big),
\end{equation}
where the bandwidth is normalized to one.

The objective is to find a power allocation such that the user sum-rate is maximized subject to power constraints:
\begin{maxi}
{\left\{P_{g}^C, \left\{P_{k,g}\right\}_{k\in \mathcal{K}_g}\right\}_{g=1}^{G}}{\sum^G_{g=1}\bigg(R_{g}^C+\sum_{k\in \mathcal{K}_g}R_{k,g}\bigg)}
{}{}
\addConstraint {\sum_{g=1}^G P_g\leq P_T}
\addConstraint{P_{k,g}\geq 0 \;\forall k\in \mathcal{K}_g\;, g=1, ..., G}
\addConstraint{P_{g}^C\geq 0\; \forall g=1, ... G.}
\label{eq:opt}
\end{maxi}
The first constraint ensures that the total allocated power over the channels meets the total power constraint, and the second and third constraints ensures positive power allocations for the private and common messages.

\section{Power Optimization for RSMA via Genetic Algorithm}

\subsection{Description of the PARGA algorithm}
We now present the proposed algorithm, used to solve (\ref{eq:opt}), dubbed Power Allocation in RSMA systems using Genetic Algorithm (PARGA).  Genetic Algorithm (GA) is a heuristic optimization technique inspired by the process of natural selection. It is used to find optimal or near-optimal solutions to complex problems, especially when the solution space is vast and traditional methods are not feasible. In GA, a population of potential solutions is evolved over generations through selection, crossover, and mutation operations, mimicking the principles of natural evolution.

In the context of power allocation for RSMA systems, we develop GA to tackle the challenging problem of determining the optimal power allocation among multiple users over the channels (\ref{eq:opt}). Finding the optimal power allocation to maximize system performance is computationally demanding, especially as the number of users increases. To address this, PARGA implements a power allocation algorithm that leverages GA optimization. This involves encoding the optimization parameters in the GA model to generate optimal power values. 

Initially, a population of potential power allocation solutions, represented as chromosomes, is randomly generated. Each chromosome encodes the power allocation for all users in the RSMA system. The fitness function is then defined to evaluate the quality of each chromosome based on system performance metrics, i.e., the sum rate. In the evolution process, the fittest chromosomes, i.e., those with the best performance, are selected to form a new population through a combination of crossover and mutation operations. Crossover involves exchanging genetic information between pairs of parent chromosomes to create offspring, while mutation introduces small random changes to the chromosomes. This mimics the natural selection process, where the most favorable traits are passed on to the next generation with occasional variations. Through multiple generations of selection, crossover, and mutation, the GA iteratively refines the population, leading to improved power allocation solutions. The algorithm continues until a termination criterion is met, i.e., reaching a maximum number of generations or achieving a satisfactory solution. 

\subsection{Pseudocode of PARGA}

The pseudocode for PARGA is provided in detail in Algorithm 1. The algorithm's parameters are specified in line 1, including the population size, mutation rate, and the elite selection rate. The input data for the algorithm is set in line 3, which includes the randomly generated power allocation blocks for all users in the RSMA system, encoded as chromosomes. The desired output is defined in line 4, which corresponds to the optimized superstring of power values.

The initialization steps are described in lines 5 to 7, where the initial strings are prepared to serve as the fitness function $\mathbf{R}$ for solving the minimization problem. Line 8 marks the beginning of the fitness function evaluation, taking into account the constraints outlined in (\ref{eq:opt}).

The algorithm iteratively improves the solution through reproduction, crossover during reproduction, and mutation of individuals from the set $\mathbf{R}$, as shown in lines 11 to 16. The superstring set $S$ is derived as the final optimized power values after reaching the maximum number of generations, as indicated in lines 22 to 26. These steps emphasize the evaluation process of the fitness function \textbf{R}, which involves the initial power value blocks \textbf{S} and adheres to the constraints specified in (\ref{eq:opt}).

\begin{figure}[h]
\centering
\label{fig_2}
\includegraphics[width=9cm]{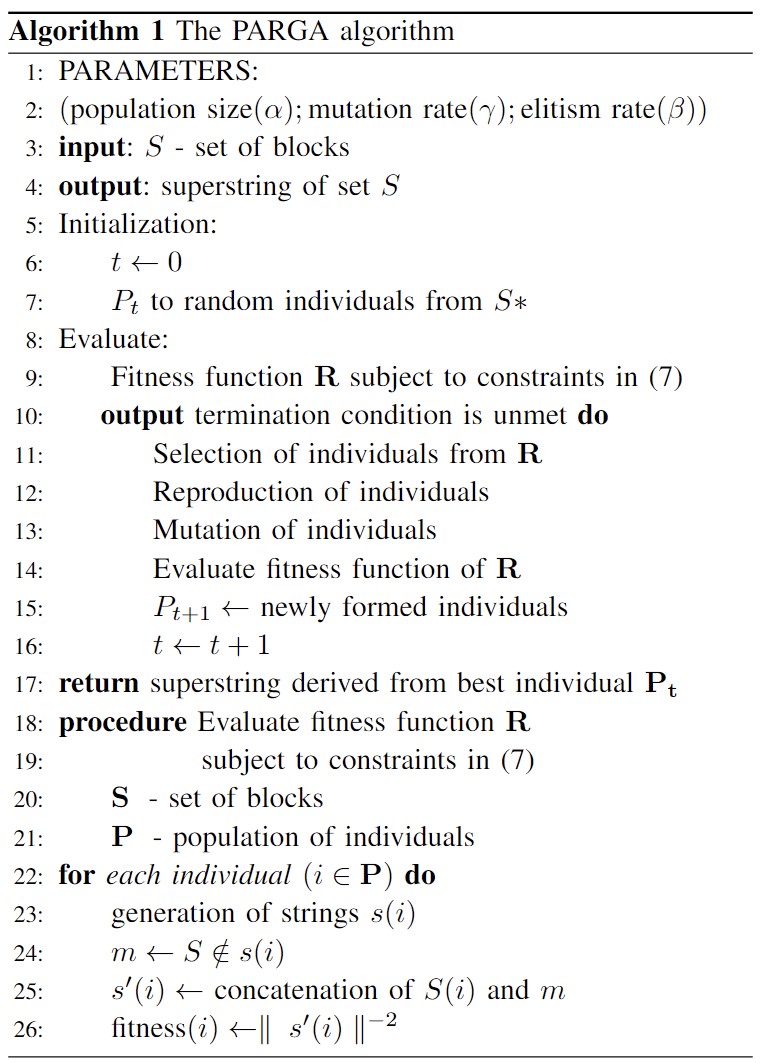}
\caption{Pseudocode of the PARGA algorithm.}
\end{figure}

\section{Simulation Results}

We simulated a common system of RSMA as introduced in \cite{mao2018rate}, consisting of a BS equipped with four transmit antennas, $N_t=4$, that serves three single-antenna users, where the the channel states of the users are given by: $\boldsymbol{h}_1 =[1,1,1,1]^H, \boldsymbol{h}_2=\gamma_1 [1,e^{j\theta_1}, e^{j2\theta_1}, e^{j3\theta_1}], \boldsymbol{h}_3=\gamma_2 [1,e^{j\theta_1}, e^{j2\theta_2}, e^{j3\theta_2}]$. The terms $\gamma_1, \gamma_2$ denote control variables of the channel gain, and $\theta_1, \theta_2$ denote control variables of the channel phase. The bandwidth is normalized to one. 

We started by evaluating the performance of a conventional fixed power allocation scheme in RSMA. We set $\theta_1=\frac{\pi}{9}, \theta_2=2\frac{\pi}{9}$. The performance is presented alongside SDMA and NOMA schemes to provide a baseline for comparison with PARGA. A proportional power allocation scheme is implemented for NOMA whereas the mean power allocation scheme aligns with the conventional power distribution methods in these schemes. For the RSMA, $50\%$ of the total power is allocated to the common message and $50\%$ is allocated to the private message decoding. We present the results for the user sum-rate in Fig. 2. The performance of the three multiple access schemes have about the same performance for SNR between $0$dB and $10$dB. The performance of RSMA improves as the SNR increases compared to SDMA and NOMA.

\begin{figure}[h]
\centering
\label{fig_2}
\includegraphics[width=9cm]{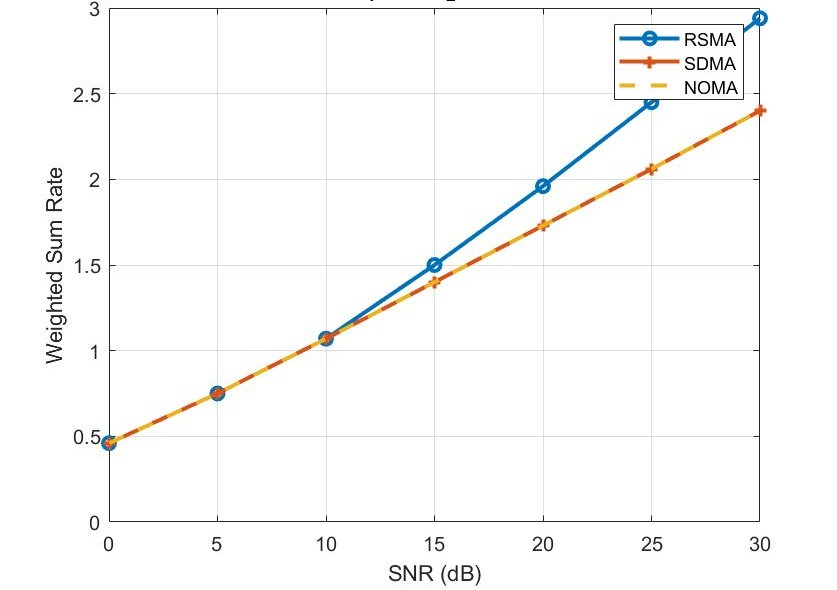}
\caption{Performance comparison of fixed power allocation at $\theta_1=\pi/9$.}
\end{figure}

Next, we compared the user sum-rate under PARGA with a fixed power allocation in RSMA systems. The results are presented in Figs. 3, 4 at $\theta_1=\pi/9$, and $\theta_1=8\pi/9$, respectively. It can be seen that PARGA significantly outperforms the fixed power allocation scheme while optimizing the power requirements in line with the constraints set in the GA simulation. These results validates the efficiency of PARGA in RSMA systems.

\begin{figure}[h]
\centering
\label{fig_6}
\includegraphics[width=9cm]{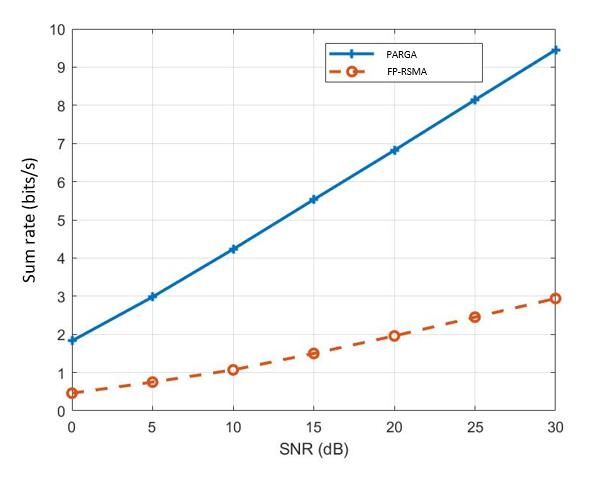}
\caption{Performance comparison between PARGA and fixed power allocation RSMA (FP-RSMA) at $\theta_1=\pi/9$.}
\end{figure}

\begin{figure}[h]
\centering
\label{fig_7}
\includegraphics[width=9cm]{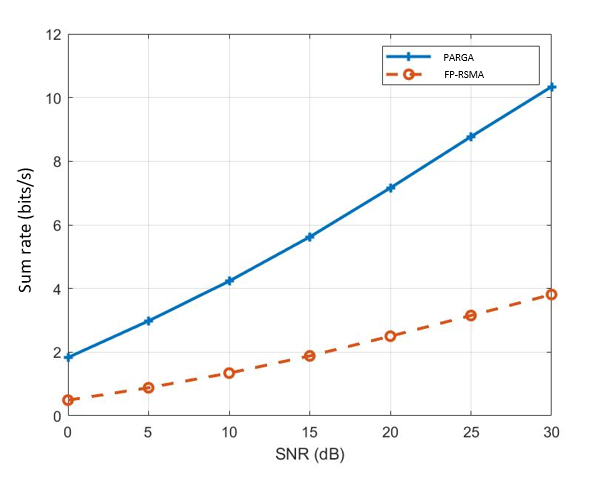}
\caption{Performance comparison between PARGA and fixed power allocation RSMA (FP-RSMA) at $\theta_1=8\pi/9$.}
\end{figure}

\section{Conclusion}

In this letter, we developed a genetic algorithm (GA)-based algorithm to optimize the power allocation between streams in RSMA systems, dubbed Power Allocation in RSMA systems using Genetic Algorithm (PARGA). PARGA aims at maximizing the user sum-rate in the network subject to a limited overall transmission power. The results revealed that PARGA significantly improves the system performance under different SNR conditions compared to a conventional fixed power allocation scheme, which makes it a promising method to implement RSMA systems. 




\end{document}